\documentclass[aps,preprint]{revtex4-1} 

\usepackage{amsmath,amssymb}    
\usepackage[dvipdfmx]{graphicx}   
\usepackage{verbatim}   
\usepackage{color}      
\usepackage{subfigure}  
\usepackage{hyperref}   
\usepackage{here}       
\usepackage{bm}         

\begin{document}

\preprint{KOBE-COSMO-16-04}

\title{MHz Gravitational Waves from Short-term Anisotropic Inflation}
\author{Asuka Ito and Jiro Soda}
\affiliation{Department of Physics, Kobe University, Kobe 657-8501, Japan}

\date{May 29, 2017}

\begin{abstract}
\noindent \hrulefill 
\begin{center} \normalsize{Abstract} \end{center}

We reveal the universality of short-term anisotropic inflation. As a demonstration, we study inflation 
with an exponential type gauge kinetic function
which is ubiquitous in models obtained by dimensional reduction from higher dimensional fundamental theory. 
It turns out that an anisotropic inflation universally takes place in the later stage of conventional inflation. 
Remarkably, we find that primordial gravitational waves  with a peak amplitude
around  $10^{-26}\sim 10^{-27}$ are copiously  produced in 
high-frequency bands 10MHz$\sim$100MHz. 
If we could detect  such gravitational waves in future,
we would be able to probe higher dimensional fundamental theory.

\noindent \hrulefill 
\end{abstract}

\maketitle 

\section{Introduction}

As is often said,  cosmology has entered into  the precision area.
 Therefore, we need to refine predictions of inflation up to the percent level and reveal qualitatively new 
 phenomena.
To this aim, we should recall that slight violation of time translation symmetry among 
 de Sitter symmetry gives rise to a tilt of the power spectrum of curvature perturbations of the order of a slow roll parameter.
The point is that the slow roll parameter characterizes deviation from the exact de Sitter spacetime.
It is interesting to observe that the non-gaussianity is also of the order of the slow roll parameter in conventional inflation.
Thus, it is legitimate to infer that the symmetry breaking is a key 
 to seek other qualitatively new phenomena of the order of the slow roll parameter.
 
Along the above line of thought, it is natural to consider the possibility of violation of rotational symmetry 
among the de Sitter symmetry,  which must lead to the statistical anisotropy in primordial fluctuations.
In fact, there were observational hints of statistical anisotropy in the CMB data~\cite{Schwarz:2015cma}. Nevertheless, 
 it was not so easy to invent a mechanism to realize the statical anisotropy in an
inflationary scenario partially because of psychological barrier due to the cosmic no-hair conjecture. 
Indeed, it had been widely believed that an anisotropy of the universe rapidly decays during inflation. 
This  is the so called the cosmic no-hair conjecture. Indeed,   
 the cosmic no-hair theorem~\cite{Wald:1983ky} where homogeneous universe with a cosmological constant is assumed 
  supported the cosmic no-hair conjecture. This is because
the inflaton potential can mimic the cosmological constant.  
Historically, there have been challenges to give a counter example to the cosmic no-hair conjecture~\cite{Ford:1989me}.
However, these models suffer from either instability, a fine tuning problem 
or a naturalness problem \cite{Himmetoglu:2008zp}. 

Eventually, an anisotropic inflationary model was found~\cite{Watanabe:2009ct},
which can be regarded as the first counter example to the cosmic no-hair conjecture. 
More precisely, it is shown that  anisotropic inflation 
occurs in the presence of a gauge kinetic function through which the gauge field is coupled to an inflaton 
and gives rise to the statical anisotropy of the order of a few percent~\cite{Gumrukcuoglu:2010yc}. 
It should be stressed that anisotropic inflation can be naturally realized in the context of supergravity.
After the discovery of anisotropic inflation, many cosmological 
predictions~\cite{Gumrukcuoglu:2010yc,WKS} and 
further extension to various models~\cite{Emami:2010rm,Kanno:2009ei,Murata:2011wv} have been discussed. 
Importantly, anisotropic inflation can be  tested with observations~\cite{Soda:2012zm,Soda:2012+}. 

 In previous works, it has been assumed that the gauge kinetic function has a specific relation to
the inflaton potential function. Actually, it is not necessary to assume the specific functional form 
for the presence of short-term  anisotropic inflation. 
Indeed, short-term anisotropic inflation occurs universally.
In this paper, we emphasize this point and study its phenomenological consequence.
As a concrete demonstration, we study inflation with the exponential type gauge kinetic function. 
As is well known, the exponential function is ubiquitous from 
the point of view of dimensional reduction from higher dimensional fundamental theory. 
As to the inflaton potential, we consider a monomial potential which can drive chaotic inflation.
It turns out that anisotropic inflation occurs in the last several e-folds. 
In fact, the gauge field starts to grow in the last stage of inflation.
We show that this universally happens for any monomial potential. 

In the conventional inflation, primordial gravitational waves are produced from vacuum quantum fluctuations.
In the present model, due to the coupling between the inflaton and the gauge field, copious
gravitational waves are produced by the classical gauge field. Since the production occurs in the last stage of the inflation,
 the frequency range of gravitational waves is typically  10MHz$\sim$100MHz.
The amplitude of gravitational waves depends on model parameters. In fact, it is easy to exceed the bound coming from the nucleosynthesis.
Hence, it is important to observationally explore  primordial gravitational waves with MHz frequencies 
from short-term anisotropic inflation. 
Although the current sensitivity of observations is too low to detect the gravitational waves~\cite{Kuroda:2015owv}, 
 future detectors~\cite{newexp} may achieve the required sensitivity. 
The recent discovery of gravitational waves~\cite{Abbott:2016blz} encourages us to pursue
the study of high frequency gravitational waves.

The paper is organized  as follows. 
In section II, we discuss the universality of short-term anisotropic inflation.
 As a demonstration, we show that short-term anisotropic inflation occurs even in a model with
 the monomial inflaton potential and the exponential type gauge kinetic function. 
We  explain how this occurs in the case of chaotic inflation in detail. 
In section III, we discuss quantization of the gauge field during anisotropic inflation
and derive mode functions for later calculations. 
In section IV, we calculate the power spectrum of the gravitational waves induced by 
 the gauge field during anisotropic inflation. 
It turns out that copious gravitational waves can be produced in the MHz frequency range. 
The final section is devoted to the conclusion.

\section{Universality of Short-term Anisotropic inflation}

In this section, first we review anisotropic inflation  \cite{Watanabe:2009ct}. 
We stress the universality of short-term anisotropic inflation. 
Then, we investigate a specific exponential type gauge kinetic function 
to reveal universality of short-term anisotropic inflation.  

We consider an action
\begin{equation}
    S=\int d^{4}x \sqrt{-g}\left[ 
                     \frac{M_{pl}^{2}}{2}R-\frac{1}{2}(\partial_{\mu}\phi)(\partial^{\mu}\phi)
                          - V(\phi) -\frac{1}{4}  f^{2}(\phi) F_{\mu\nu}F^{\mu\nu}
                              \right] \ ,  \label{action}
\end{equation}
where $M_{pl}$ represents the reduced Plank mass, $g$ is the determinant of the metric $g_{\mu\nu}$ and 
$R$ is the Ricci scalar. Here, we introduced a gauge field $A_\mu$ with the field strength
 $F_{\mu\nu}=\partial_{\mu}A_{\nu}-\partial_{\nu}A_{\mu}$ and
  the gauge field is coupled to an inflaton $\phi$ through a gauge kinetic function  $f(\phi)$. 
For the moment, we do not specify a potential function for the inflaton $V(\phi)$.

Let us discuss the cosmological homogeneous dynamics. The homogeneity imposes $\phi =\phi(t)$.
For the gauge field, we can take an ansatz $A_{\mu}=(0,v_{A}(t),0,0)$ without loss of generality. 
In the presence of the non-trivial gauge field, we can not take isotropic ansatz for the metric. Instead,  we
take the anisotropic ansatz 
\begin{equation}
    ds^{2}=-dt^{2}+e^{2\alpha(t)}\left[ e^{-4\sigma(t)}dx^{2}+e^{2\sigma(t)}(dy^{2}+dz^{2}) \right] \ . 
\label{eq2}
\end{equation}
Now, we can derive equations of motion. It is easy to solve the equation for the gauge field as
\begin{eqnarray}
    \dot{v}_{A}=p_{A}f^{-2}(\phi)e^{-\alpha-4\sigma} \ . 
   \label{integral}
\end{eqnarray}
where $p_{_{A}}$ is an integration constant.
Thus, the field equation for the inflaton  reads
\begin{eqnarray}
    \ddot{\phi}&=&-3\dot{\alpha}\dot{\phi}-V_{,\phi}+
                p_{A}^{2}f_{,\phi} f^{-3}(\phi) e^{-4\alpha-4\sigma}
                  \ , \label{infeq}
\end{eqnarray}
From the Einstein equation,  we obtain the hamiltonian constraint 
\begin{eqnarray}
    \dot{\alpha}^{2}=\dot{\sigma}^{2}+\frac{1}{3M_{pl}^{2}}\left[\frac{1}{2}\dot{\phi}^{2}+V(\phi)+
                     \frac{1}{2}p_{A}^{2}f^{-2}(\phi) e^{-4\alpha-4\sigma} \right] \  ,
                   \label{hamiconst}
\end{eqnarray}
and the rest of Einstein equations
\begin{eqnarray}
    \ddot{\alpha}&=&-3\dot{\alpha}^{2}+\frac{1}{M_{pl}^{2}}V(\phi)+
                   \frac{1}{6M_{pl}^{2}}p_{A}^{2}f^{-2}(\phi)e^{-4\alpha-4\sigma} 
                   \ ,\label{eq4}\\
    \ddot{\sigma}&=&-3\dot{\alpha}\dot{\sigma}+
                  \frac{1}{3M_{pl}^{2}}p_{A}^{2}f^{-2}(\phi)e^{-4\alpha-4\sigma}
                   \  .    \label{eq5}
\end{eqnarray}

To understand how anisotropic inflation occurs, 
it is convenient to express the gauge kinetic function in terms of the scale factor $a(t)$ as 
\begin{equation}
    f(\phi)\propto a^{-n}(t)=e^{-n\alpha}\ . \label{coupfun}
\end{equation}
From  Eq.(\ref{hamiconst}), one can see that the energy density of the gauge field grows if $n>2$ . 
It decays for $n<2$. In the case of $n=2$, it is almost constant. 
Note that $\sigma$ is always negligible compared with $\alpha$. 
In the regime where the inflaton slow rolls and the back reaction from the gauge field is negligible, 
  Eqs.(\ref{infeq}) and  (\ref{hamiconst}) can be reduced to 
\begin{eqnarray}
   3\dot{\alpha}\dot{\phi}&\simeq&-V_{,\phi} \ ,\\
   \dot{\alpha}^{2}&\simeq&\frac{1}{3M_{pl}^{2}}V(\phi)  \ .
\end{eqnarray}
These equations lead to 
\begin{equation}
    \frac{d\alpha}{d\phi}=-\frac{1}{M_{pl}^{2}}\frac{V}{V_{,\phi}} \ .
\end{equation}
Integrating it,  we obtain
\begin{equation}
    \alpha = -\int \frac{1}{M_{pl}^{2}}\frac{V}{V_{,\phi}}d\phi+D \label{alpha2} \ ,
\end{equation}
where $D$ is a constant of  integration. 
Thus, we  find  functional form of the gauge kinetic function (\ref{coupfun}) as 
\begin{equation}
    f(\phi)=e^{\frac{n}{M_{pl}^{2}} \int \frac{V}{V_{,\phi}} d\phi }\ . \label{coupfun2}
\end{equation}
Although the expression (\ref{coupfun2}) is useful for discussing concrete models of anisotropic inflation,
it is too restrictive. In fact, it requires the gauge field is always active during inflation.
 However, it is interesting to consider the possibility that the gauge field becomes relevant for a certain period during inflation.
Actually, as stressed in \cite{Soda:2012zm}, the condition for the occurrence of anisotropic inflation reads
\begin{eqnarray}
 M_{pl}^2 \frac{f_{,\phi}}{f}\frac{V_{,\phi}}{V} > 2  \ . \label{general_cond}
\end{eqnarray}
This condition determines the region where short-term anisotropic inflation occurs.
 In the extreme case (\ref{coupfun2}), 
 the gauge field is sustained in the whole period of inflation. 

From now on, as a concrete demonstration, we consider the gauge kinetic  function 
\begin{equation}
    f(\phi)=e^{\frac{c}{M_{pl}} \phi} \label{exp}\ ,
\end{equation}
where $c$ is a coupling constant. 
It should be noted that such an exponential type gauge kinetic function is ubiquitous in models obtained by  dimensional reduction
 from higher dimensional fundamental theory such as superstring theory. 
 Apparently, from the criterion (\ref{coupfun2}),   anisotropic inflation does not happen.
In fact,  however, for a certain constant $c$, the gauge field can grow. 
We will show it explicitly. 
For  the gauge field to grow, the decreasing rate of the gauge kinetic function must be larger than 
that of  (\ref{coupfun2}) with  $n=2$.  Note that  Eq.(\ref{coupfun2}) is a decreasing function.
When we choose the functional form (\ref{exp}), this condition is given by 
\begin{eqnarray}
      \qquad \frac{d}{dt}\left[ \exp\left (\frac{c}{M_{_{pl}}}\phi-\frac{2}{M^{2}_{pl}} 
                                 \int \frac{V}{V_{,\phi}} d\phi \right) \right] < 0 \ . 
\end{eqnarray}
Of course, this does not hold in general. Rather, it constrains the region where anisotropic inflation is possible.
Indeed, we have the condition
\begin{equation}
    c\ > \frac{2}{M_{pl}}\frac{V}{V_{,\phi}} \ , \label{condition}
\end{equation}
where we used the fact that the derivative of the inflaton with respect to time is negative.
Notice that this is nothing but the condition (\ref{general_cond}).  
If  the inequality (\ref{condition}) is satisfied, the gauge field grows.
To be more precise,  let us consider a monomial potential 
\begin{equation}
    V(\phi)\propto \phi^{l}\label{powerlaw}  \ ,
\end{equation}
where $l$ is a real parameter.   Then it becomes 
\begin{equation}
    c > \frac{2}{l} \frac{\phi}{M_{pl}} \label{condition2}\ .
\end{equation}
Thus, in the region where the condition  (\ref{condition2}) is satisfied, we can realize anisotropic inflation. 
For example, for the potential  (\ref{powerlaw}), the  inflaton takes the value  
$\phi \sim \mathcal{O} \left(10 M_{pl}\right)$ around CMB scales
provided the number of e-foldings  $50\sim 60$, and the inflation ends around  $\phi \sim \mathcal{O} \left(1 M_{pl}\right)$.
Hence,  if $c > \mathcal{O}(1)$ we can see the growth of the  gauge field in the late stage of inflation. 
\ 

In Eq.(\ref{coupfun}), we assumed constancy of  $n$. Then, we obtained the functional form (\ref{coupfun2}).
Since we took  the gauge coupling function (\ref{exp}) different from the function (\ref{coupfun2}), 
$n$ should be time dependent. 
Let us try to express the gauge kinetic function (\ref{exp}) in the  form (\ref{coupfun}) 
with the time dependent $n(t)$: 
\begin{eqnarray}
    e^{\frac{c}{M_{pl}}\phi} \propto a^{-n(t)}=e^{-n(t)H_{I}t} \ , \label{view}
\end{eqnarray}
where $H_{I}$ is the Hubble constant during inflation. 
This gives rise to the following relation
\begin{equation}
    \frac{c}{M_{pl}}\dot{\phi} = -\dot{n}H_{I}t-nH_{I} \ \label{ndot} \ .
\end{equation}
Here, we assume the first term of the right-hand side of Eq.(\ref{ndot}) is negligible  compared with the second term 
in a time scale of expansion $H_{I}^{-1}$. 
Then, we get 
\begin{equation}
    n=-\frac{c}{M_{pl}H_{I}}\dot{\phi} \ .
\end{equation}
From this relation, one can take a derivative of $n(t)$ with respect to the time 
\begin{equation}
    \frac{\dot{n}}{H_{I}n}= \frac{\ddot{\phi}}{H_{I}\dot\phi}\ .
\end{equation}
If the slow roll condition is satisfied, it is much smaller than 1. 
Therefore we can treat $n(t)$ as an almost constant quantity during slow roll inflation.
\begin{figure}[!h]
\begin{center}
\includegraphics[width=9cm]{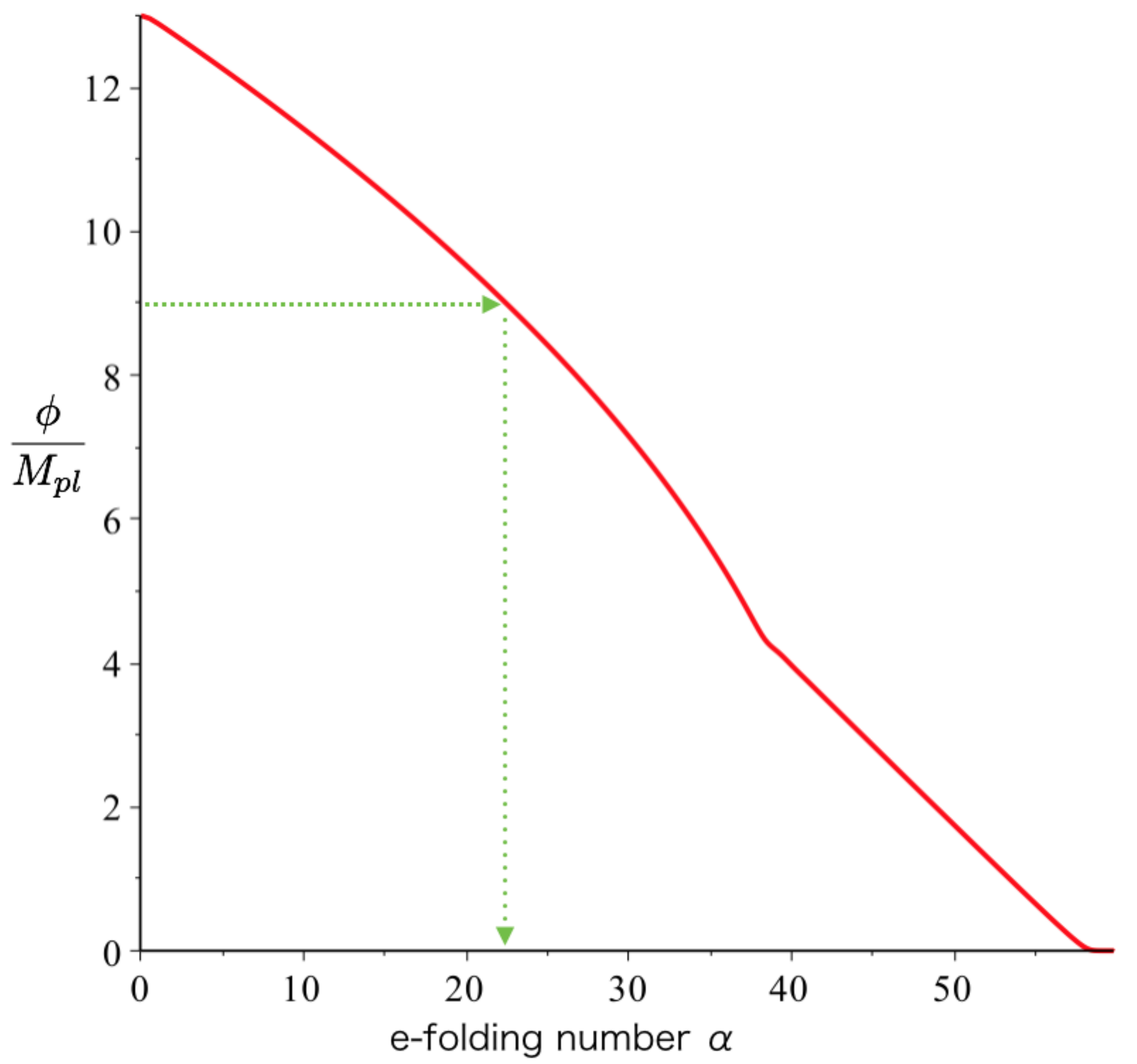}
\end{center}
\caption{The evolution of the inflaton field is depicted. The green vector indicates the e-folding number 
corresponding to  $\frac{\phi}{M_{pl}}=9(=c)$. 
The transition to the second inflationary phase occurs around $\alpha \simeq 38$.}
\label{phifig}
\end{figure}

\begin{figure}[!h]
\begin{center}
\includegraphics[width=9cm]{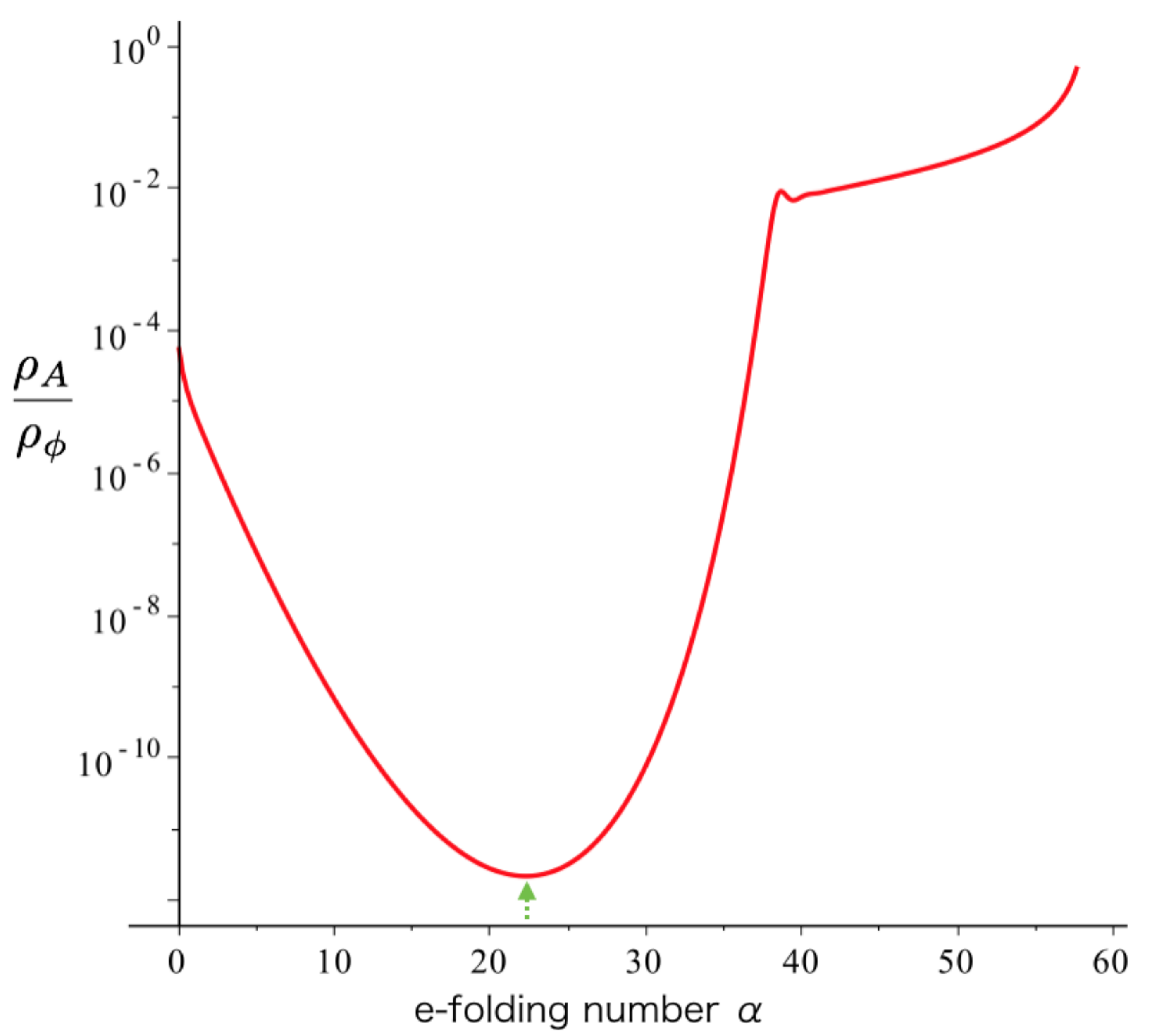}
\end{center}
\caption{The ratio of the energy density of the gauge field to that of the inflaton field is plotted. 
The green vector indicates the moment when the inflaton 
$\frac{\phi}{M_{pl}}=9(=c)$. 
After  $\alpha \simeq 38$, the ratio becomes nearly a constant.}
\label{anisotropy}
\end{figure}

\begin{figure}[!h]
\begin{center}
\includegraphics[width=9cm]{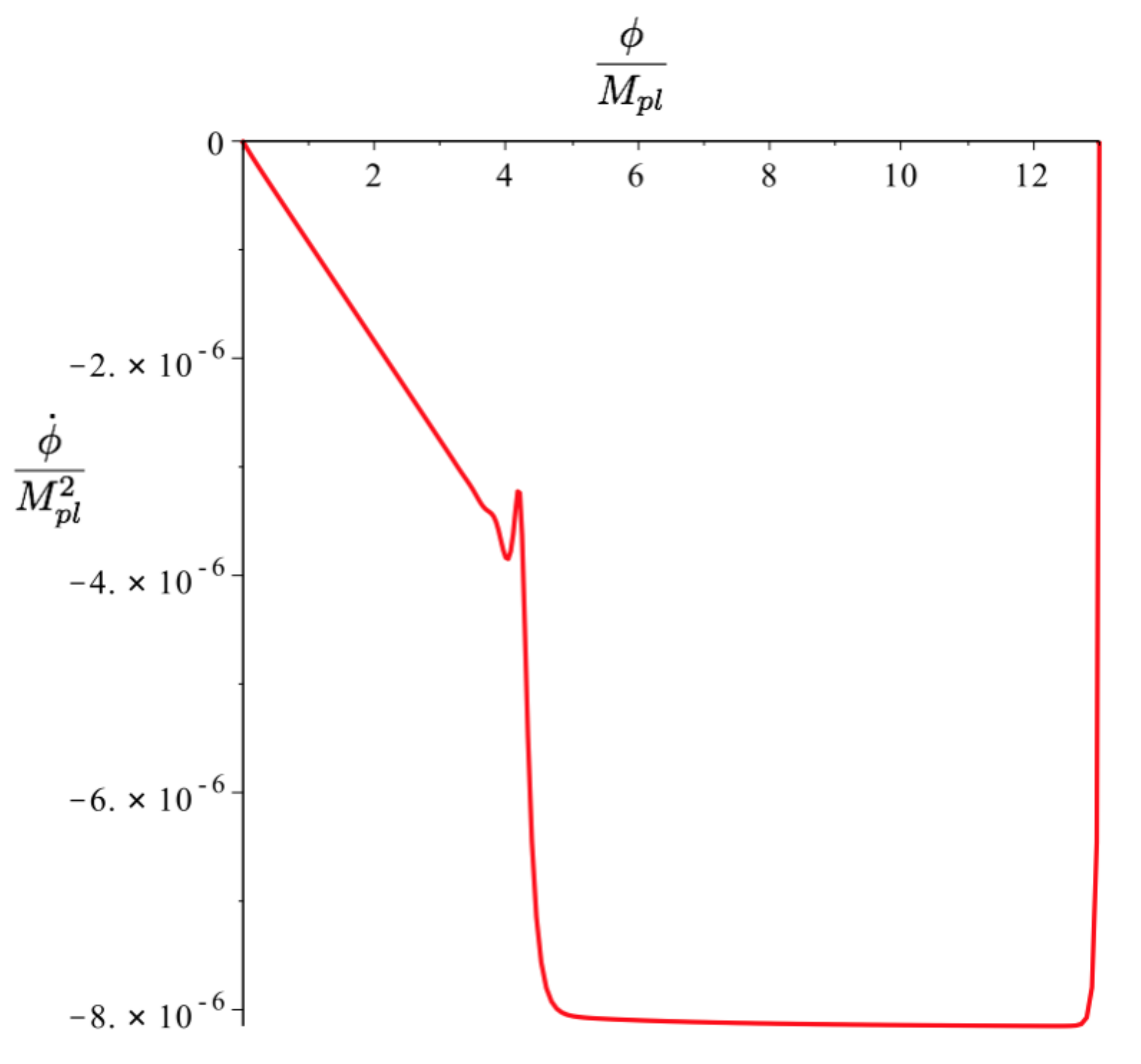}
\end{center}
\caption{The trajectory of the inflaton field in the phase space is depicted. 
We took the initial conditions  $\phi_{0}=13M_{pl}$ and $\dot{\phi}_{0}=0$. 
The transition to the second inflationary phase occurs around $\phi\simeq 4 M_{pl}$.}
\label{phasespace}
\end{figure}

Now let us examine a concrete example of short-term anisotropic inflation with the exponential gauge kinetic function
and the potential $V=\frac{1}{2}m^{2}\phi^{2}$.  Here, $m$ is the mass of the inflaton.
In this case,  Eqs.(\ref{infeq})-(\ref{eq5}) lead to 
\begin{eqnarray}
    \dot{\alpha}^{2}&=&\dot{\sigma}^{2}+\frac{1}{3M_{pl}^{2}}\left[\frac{1}{2}\dot{\phi}^{2}+\frac{1}{2}m^{2}\phi^{2}+
                     \frac{1}{2}p_{A}^{2}e^{-2\frac{c}{M_{pl}}\phi} e^{-4\alpha-4\sigma} \right]\ ,\quad 
                   \label{hamiconst2}\\
    \ddot{\alpha}&=&-3\dot{\alpha}^{2}+\frac{1}{2 M_{p}^{2}}m^{2}\phi^{2}+
                   \frac{1}{6M_{p}^{2}}p_{A}^{2}e^{-2\frac{c}{M_{pl}}\phi} e^{-4\alpha-4\sigma} 
                   \ ,\label{alpha}\\
    \ddot{\sigma}&=&-3\dot{\alpha}\dot{\sigma}+
                  \frac{1}{3M_{p}^{2}}p_{A}^{2}e^{-2\frac{c}{M_{pl}}\phi}e^{-4\alpha-4\sigma}
                   \ ,\label{sigma}\\
    \ddot{\phi}&=&-3\dot{\alpha}\dot{\phi}  -  m^{2} \phi +
                p_{A}^{2} \frac{c}{M_{pl}} e^{-2\frac{c}{M_{pl}}\phi} e^{-4\alpha-4\sigma}
                  \ . \label{infeq2}
\end{eqnarray}
We solve the above equations numerically.
The results are shown in Figs.\ref{phifig}-\ref{phasespace} . 
In the calculations, we have set the mass of inflaton $m=10^{-5}M_{pl}$ , 
the initial condition of the inflaton $\phi_{0}=13M_{pl}$ , and $c=9$. 
In Fig.\ref{phifig}, we plotted the evolution of the inflaton. We can see a transition around at $\alpha =38$.  
From Fig.\ref{anisotropy},  taking a look at the green dotted arrow, one can find that 
the gauge field start to grow at around  $\alpha =22$ and eventually enter into the attractor around at $\alpha=38$. 
From Fig.\ref{phifig}, the number of e-folds  $\alpha =22$ corresponds to  $\phi=9M_{pl}$. 
This is a consequence of the inequality (\ref{condition2}). 
This growth of  the gauge field made the transition of the inflation to the second inflationary phase around at $\alpha\simeq 38$ . 
In the second phase, the third term of the right-hand side of  Eq.(\ref{infeq2}) can not be negligible, namely, 
$p_{A}^{2} \frac{c}{M_{pl}} e^{-2\frac{c}{M_{pl}}\phi} e^{-4\alpha-4\sigma}\sim V_{,\phi}$ . 
Then the ratio of the energy density of the gauge field to that of the inflaton field during the second 
inflationary phase is 
\begin{equation}
    \mathcal{R}\equiv \frac{\frac{1}{2}p_{A}^{2}e^{-2\frac{c}{M_{pl}}\phi} e^{-4\alpha-4\sigma}}{V(\phi)} 
    \sim \frac{M_{pl}}{2c}\frac{V_{,\phi}}{V} \ .
\end{equation}
In the present case,  we have
\begin{equation}
    \mathcal{R} \sim \frac{M_{pl}}{c\phi} \label{ratio} \ .
\end{equation}
Thus, the energy density of the gauge field is negligible in  Eq.(\ref{hamiconst2}) even when 
the gauge field is relevant in  Eq.(\ref{infeq2}) during the second inflationary phase. 
Therefore, from  Eqs.(\ref{hamiconst2}) and (\ref{sigma}), we can get 
\begin{eqnarray}
    \dot{\alpha}^{2}&\simeq&\frac{1}{3M_{pl}^{2}} V\ ,\quad 
                   \label{hamiconst3}\\
    3\dot{\alpha}\dot{\sigma}&\simeq&
                  \frac{1}{3M_{p}^{2}}p_{A}^{2}e^{-2\frac{c}{M_{pl}}\phi}e^{-4\alpha-4\sigma}
                   \ ,\label{sigma3}
\end{eqnarray}
where we have assumed $\ddot{\sigma}\ll 3\dot{\alpha}\dot{\sigma}$.
From  Eqs.(\ref{hamiconst3}) and (\ref{sigma3}), we obtain the relation between the anisotropy 
$\frac{\dot{\sigma}}{\dot{\alpha}}$
and the ratio  $\mathcal{R}$ as follows
\begin{eqnarray}
    \frac{\dot{\sigma}}{\dot{\alpha}}  \simeq
      \frac{1}{3} \frac{p_{A}^{2}e^{-2\frac{c}{M_{pl}}\phi}e^{-4\alpha-4\sigma}}{V} 
    \sim \frac{M_{pl}}{3c}\frac{V_{,\phi}}{V} 
    \sim \frac{2}{3}\mathcal{R} \ .
\end{eqnarray}
We note that  Eq.(\ref{ratio}) holds during the attractor. 
Considering the effective potential of the inflaton field, it turns out that the attractor like 
behavior means the inflaton falls the trough of the effective potential. 
Numerical result Fig.\ref{phasespace}  tells us that  the inflaton moves with the trough. 
In the second inflationary phase, the rolling of the inflaton is governed by the effective potential 
corrected by the gauge field, so that  the velocity of the inflaton  decreases gradually. 
This result is different from that of conventional anisotropic inflation with the gauge kinetic function 
(\ref{coupfun2}) \cite{Watanabe:2009ct}. 
From the view of  Eq.(\ref{view}), we can understand that 
the difference comes from the growth of the power $n$. 

In the above example, we took $c=9$. In this case, at some point the gauge field starts to grow and continues to grow for about 15 e-folds,   
 and enter into the attractor phase of short-term anisotropic inflation which persists about 20 e-folds. 
For a more moderate parameter $c$, the duration of short-term anisotropic inflation becomes more short.
In the extreme case, there would be no attractor phase and the period where the gauge field is relevant would be
about 4 e-folds. It is the case that we are focusing on in the rest of this paper. 

\section{Quantum gauge field during anisotropic inflation}
In this section, we derive the mode function of the gauge field which will be used for calculations
of gravitational waves in the next section. 

Neglecting the back reaction from the gauge field, we consider the isotropic inflationary background 
\begin{equation}
    ds^{2}=a(\tau)^{2} \left[ -d \tau^{2} +dx^{2}+dy^{2}+dz^{2} \right] \label{metric} \ ,
\end{equation}
where $\tau$ is the conformal time.\ 
One can expand the gauge field in Fourier space as 
\begin{equation}
    \vec{A}(\tau,\vec{x})=\int \frac{d^{3}k}{(2\pi)^{3/2}} 
                         \bm{A}_{\bm{k}}(\tau) e^{i \bm{k} \cdot \bm{x}}  \ .
\end{equation}
Then the part for the gauge field in the action (\ref{action}) can be rewritten as 
\begin{equation}
    S_{gauge}=\frac{1}{2} \int d\tau d^{3}k f^{2}(\phi) 
    \left[ \bm{A}'_{\bm{k}}\bm{A}'_{-\bm{k}}-
    k^{2}\bm{A}_{\bm{k}}\bm{A}_{\bm{-k}} \right] \ ,\label{fac}
\end{equation}
where a prime represents a derivative with respect to the conformal time. 
The Fourier mode of the vector potential can be promoted into the operator and
 expanded by the creation and annihilation operators satisfying 
$\left[ \hat{a}^{(\sigma)}_{\bm{k}},\hat{a}^{(\rho)\dagger}_{\bm{k}'} \right]=
\delta_{\sigma\rho}\delta(\bm{k}-\bm{k}')$ as 
\begin{equation}
    \hat{\bm{A}}_{\bm{k}}(\tau)=\sum_{\sigma=+,-} \vec{e}^{(\sigma)}(\hat{k}) \left[ 
                                  U_{k}(\tau)\hat{a}^{(\sigma)}_{\bm{k}}+
                                  U_{k}^{*}(\tau)\hat{a}^{(\sigma)\dagger}_{-\bm{k}} \right]\ ,
\end{equation}
where $\sigma$ represents the two polarization degrees of the freedom of the vector potential. 
The circular polarization vector $\vec{e}^{(\sigma)}$  satisfies the relations
\begin{eqnarray}
    \vec{k}\cdot \vec{e}^{(\pm)}(\hat{k})&=& 0 \quad ,\nonumber \\
    \vec{k}\times \vec{e}^{(\pm)}(\hat{k})&=&\mp i k \vec{e}^{(\pm)}(\hat{k})\quad ,\nonumber \\
    \left(\vec{e}^{(\pm)}(\hat{k})\right)^{*}=\vec{e}^{(\pm)}(-\hat{k})\ ,&& \  
                \left(\vec{e}^{(\sigma)}(\hat{k})\right)^{*} \cdot \vec{e}^{(\rho)}(\hat{k})
                  =\delta_{\sigma\rho} \quad .
\end{eqnarray}
The mode function obeys the equation  derived from the action (\ref{fac})
\begin{equation}
    U''_{k}+2\frac{f'}{f}U'_{k}+k^{2}U_{k}=0 \label{modef} \ .
\end{equation}
Using the new variable $u_{k}\equiv f U_{k}$ , we get 
\begin{equation}
    u_{k}''+\left( k^{2}-\frac{f''}{f}\right) u_{k}=0 \label{mode}\ .
\end{equation}
Hereafter we consider the exponential type functional form (\ref{exp}). 
Then, we get 
\begin{equation}
    \frac{f''(\tau)}{f(\tau)}=\frac{c \phi''(\tau)}{M_{pl}}+
                              \frac{c^{2}\phi'^{2}(\tau)}{M_{pl}^{2}} \ .
                              \label{40}
\end{equation}
Using the cosmic time $t$, the derivative of the $\phi$ with respect to $\tau$ is rewritten as 
\begin{equation}
    \phi'(\tau)=a(\tau) \dot{\phi}(t) \label{phidash}\ ,
\end{equation}
where an overdot denotes a derivative with respect to the cosmic time.
Assuming the slow roll inflation, we obtain 
\begin{eqnarray}
    a(\tau)&\simeq& -\frac{1}{H_{I}\tau} \label{scalefactor}\ , \\
    \dot{\phi}&\simeq& - \sqrt{2\epsilon} M_{pl} H_{I} \label{phidot} \ ,
\end{eqnarray}
where $H_{I}$ is the Hubble constant during inflation and $\epsilon \equiv -\frac{\dot{H}_{I}}{H_{I}^{2}}$ is 
the slow roll parameter.
From these equations,  Eq.(\ref{phidash}) becomes
\begin{equation}
    \phi'(\tau)=\frac{\sqrt{2\epsilon}M_{pl}}{\tau} \ .
\end{equation}
Since $\epsilon$ is almost  constant, we can ignore the derivative of $\epsilon$. 
Hence, we get 
\begin{equation}
    \phi''(\tau)\simeq -\frac{\sqrt{2\epsilon}M_{pl}}{\tau^{2}}\ .
\end{equation}
Thus, Eq.(\ref{40}) reduces to
\begin{equation}
    \frac{f''(\tau)}{f(\tau)}\simeq \frac{-c\sqrt{2\epsilon}+2c^{2}\epsilon}{\tau^{2}} \label{nanka}\ .
\end{equation}
Substituting  Eq.(\ref{nanka}) into  Eq.(\ref{mode}) and solving it with the Bunch-Davies initial 
condition, we get  the mode function
\begin{equation}
    u_{k}(\tau)=\frac{1}{\sqrt{2k}}\sqrt{\frac{-k\tau \pi}{2}}H_{\nu-\frac{1}{2}}^{(1)}(z) \label{modesol}\ ,
\end{equation}
where $H_{\gamma}^{(1)}(x)$ is the Hankel function of the first kind, 
$z\equiv -k\tau$ and 
\begin{equation}
    \nu\equiv c\sqrt{2\epsilon}\label{nu}\ .
\end{equation}
Now, we define the electric and magnetic fields  as
\begin{equation}
    \vec{E}(\tau, \bm{x})\equiv -\frac{f}{a^{2}}\partial_{\tau} \vec{A}(\tau,\bm{x})\ , \quad
    \vec{B}(\tau, \bm{x})\equiv \frac{f}{a^{2}}\left( \nabla\times \vec{A}(\tau,\bm{x}) \right) \ \label{dif}.
\end{equation}
They can be expanded in Fourier space as 
\begin{eqnarray}
    \vec{E}(\tau,\bm{x})&=&\int \frac{d^{3}k}{(2\pi)^{3/2}} \hat{\bm{E}}_{\bm{k}}(\tau)
                          e^{i\bm{k}\cdot\bm{x}} \ \label{E},\\
    \vec{B}(\tau,\bm{x})&=&\int \frac{d^{3}k}{(2\pi)^{3/2}} \hat{\bm{B}}_{\bm{k}}(\tau)
                          e^{i\bm{k}\cdot\bm{x}} \ ,\label{BB}
\end{eqnarray}
where 
\begin{eqnarray}
    \hat{\bm{E}}_{\bm{k}}(\tau)&=&\sum_{\sigma=+,-}\vec{e}^{(\sigma)}(\hat{k})
                             \left[ \mathcal{E}_{k}\hat{a}^{(\sigma)}_{\bm{k}}+
                              \mathcal{E}^{*}_{k}\hat{a}^{(\sigma)\dagger}_{-\bm{k}}\right] \ ,\\
    \hat{\bm{B}}_{\bm{k}}(\tau)&=&\sum_{\sigma=+,-}\sigma \vec{e}^{(\sigma)}(\hat{k})
                              \left[ \mathcal{B}_{k}\hat{a}^{(\sigma)}_{\bm{k}}+
                              \mathcal{B}^{*}_{k}\hat{a}^{(\sigma)\dagger}_{-\bm{k}}\right] \ . \label{B}
\end{eqnarray}
Here, we defined
\begin{eqnarray}
    \mathcal{E}_{k}(\tau)=-\frac{f}{a^{2}}\partial_{\tau}U_{k}(\tau) \ ,\quad 
    \mathcal{B}_{k}(\tau)=\frac{f}{a^{2}} k U_{k}(\tau) \ .
\end{eqnarray}
Using the mode function (\ref{modesol}), we obtain 
\begin{equation}
    \mathcal{E}_{k}(\tau)=\sqrt{\frac{\pi}{2}} k^{-3/2} \left( H_{I}z \right)^{2} \sqrt{\frac{z}{2}} 
                       H^{(1)}_{\nu+\frac{1}{2}}(z)\ , \quad 
    \mathcal{B}_{k}(\tau)=\sqrt{\frac{\pi}{2}} k^{-3/2} \left( H_{I}z \right)^{2} \sqrt{\frac{z}{2}} 
                       H^{(1)}_{\nu-\frac{1}{2}}(z)\ .\label{elemag}
\end{equation}
In the super horizon limit $|z|\rightarrow 0$,  we can use the approximation
\begin{equation}
    H^{(1)}_{\nu}(z)\simeq 
                                 - \frac{i\Gamma(\nu)}{\pi}\left( \frac{2}{z}\right)^{\nu} 
                             \label{hankel}\ .
\end{equation}
Now, we consider only the case $\nu > 2$.
Then the electric field and the magnetic field are
\begin{eqnarray}
    \mathcal{E}_{k}(\tau)&=&\frac{i\Gamma(\nu+\frac{1}{2})}{\sqrt{\pi}}H^{2}_{I} \left( \frac{2}{k} \right)^{3/2} 
                         \left( \frac{2}{z} \right)^{\nu-2}, \label{ele}\\
    \mathcal{B}_{k}(\tau)&=&-\frac{i\Gamma(\nu-\frac{1}{2})}{\sqrt{\pi}}H^{2}_{I} \left( \frac{2}{k} \right)^{3/2} 
                         \left( \frac{2}{z} \right)^{\nu-3}\ \label{mag}.
\end{eqnarray}
We can find that the electric field always dominates over the magnetic field and only when $\nu > 2$, 
it grows in the super horizon regime.
Substituting the slow roll parameter by the inflaton potential 
$\epsilon \simeq \frac{M^{2}_{pl}}{2}\left( \frac{V_{,\phi}}{V} \right)^{2}$ into Eq.(\ref{nu}), we have 
\begin{equation}
    \nu \simeq c M_{pl} \left( \frac{V_{,\phi}}{V} \right)\label{nupote}\ .
\end{equation}
Then the condition $\nu > 2$ yields
\begin{equation}
    c > \frac{2}{M_{pl}}\frac{V}{V_{,\phi}} \ .
\end{equation}
This coincides with  the condition  (\ref{condition}).

\section{MHz Gravitational waves induced by  gauge field}
In this section, we consider short-term anisotropic inflation with $ c\sim 3$ for which typically  there is no an attractor phase.
 We show that copious gravitational waves can be produced by the gauge field during 
the phase approaching the attractor. 
The analysis  is quite similar to that used in a different context~\cite{Barnaby:2012tk}. 

\begin{figure}[h]
\begin{center}
\includegraphics[width=11cm]{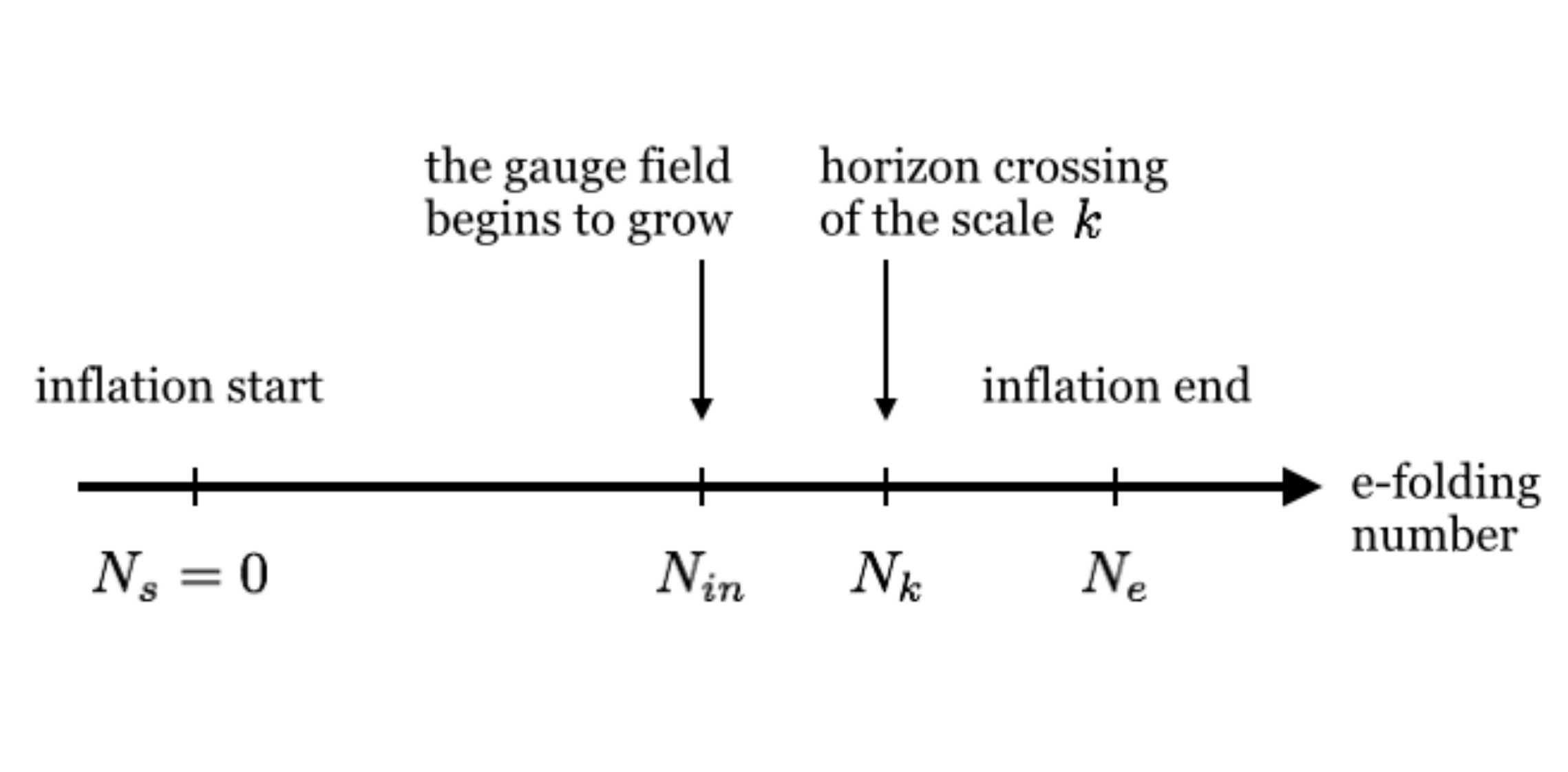}
\end{center}
\caption{We have defined each  relevant moment by the number of e-foldings.}
\label{fig1}
\end{figure}

Since we consider the situation where the gauge field is negligible in background equations, 
the energy momentum tensor, namely the nonlinear part, of the gauge field becomes the source of gravitational waves. 
One can get the tensor sector of the action (\ref{action}) as 
\begin{equation}
    S_{GW}=\int d\tau d^3x \left[\  \frac{M^{2}_{pl}}{8}a^{2} 
       \left( h'_{ij}h'^{ij}-\partial_{k} h_{ij} \partial_{k} h^{ij} \right)   
            +\frac{1}{2}a^4\left( E_{i}E_{j}+B_{i}B_{j} \right) h^{ij} \ \right] \ \label{action5},
\end{equation}
where $h_{ij}$ is a  transverse traceless tensor and we used the definition of the 
electric and magnetic fields (\ref{dif}). 
The tensor fluctuation can be expanded in Fourier space as 
\begin{eqnarray}
    h_{ij}&=&\sum_{\sigma=+,-} \int \frac{d^3k}{(2\pi)^{3/2}} 
           \hat{h}_{\bm{k}}^{(\sigma)} e^{i \bm{k}{\bm{x}}} \Pi ^{(\sigma)}_{ij} \ , \label{hfourier}\\
    \hat{h}_{\bm{k}}^{(\sigma)}(\tau)&=&V_{k}(\tau)\hat{a}_{\bm{k}}^{(\sigma)}+
                                  V_{k}^{*}(\tau)\hat{a}_{-\bm{k}}^{(\sigma) \dagger}\ ,
\end{eqnarray}
where $\Pi ^{(\sigma)}_{ij}$ 
are polarization tensors constructed by circular polarization vectors as 
$\Pi ^{(\sigma)}_{ij}\equiv e_{i}^{(\sigma)}e_{j}^{(\sigma)}$ and 
we have used creation and annihilation operators. 
Substituting  Eqs.(\ref{E})-(\ref{B}) and (\ref{hfourier}) into  Eq.(\ref{action5}), we obtain 
\begin{eqnarray}
    S_{GW}=\sum_{\sigma=+,-}&&\int d\tau d^3k \biggl[\ \frac{M^{2}_{pl}}{4}a^{2} 
    \left( \hat{h}'^{(\sigma)}_{\bm{k}}\hat{h}'^{(\sigma)}_{\bm{-k}}-
    k^2 \hat{h}^{(\sigma)}_{\bm{k}}\hat{h}^{(\sigma)}_{\bm{-k}} \right)  \nonumber \\
   -  \frac{a^4}{2}&&\int \frac{d^{3}p}{(2\pi)^{3/2}}
    \left( \hat{E}_{i,\bm{p}}\hat{E}_{j,\bm{k}-\bm{p}}+\hat{B}_{i,\bm{p}}\hat{B}_{j,\bm{k}-\bm{p}} \right)
     e^{*(\sigma)}_{i}(\hat{k})e^{*(\sigma)}_{j}(\hat{k}) 
     \ \hat{h}_{\bm{-k}}^{(\sigma)} \ \biggr] \ .
\end{eqnarray}
Using the normalized variable $v_{k}\equiv \frac{M_{pl}}{2}aV_{k}$, 
we can get the equation for the mode function of the gravitational waves as 
\begin{equation}
    v_{k}''(\tau)+\left( k^{2}-\frac{2}{\tau^{2}} \right) v_{k}(\tau) 
                   = S^{(\sigma)}(\tau,\bm{k}) \label{hmode}\ ,
\end{equation}
where the source term is defined by 
\begin{equation}
    S^{(\sigma)}(\tau,\bm{k}) = -\frac{a^{3}}{M_{pl}} \int \frac{d^{3}p}{(2\pi)^{3/2}} 
                                    \left( \hat{E}_{i,\bm{p}}\hat{E}_{j,\bm{k}-\bm{p}}+\hat{B}_{i,\bm{p}}\hat{B}_{j,\bm{k}-\bm{p}}\right)
                            e^{*(\sigma)}_{i}(\hat{k})e^{*(\sigma)}_{j}(\hat{k}) \ .
\end{equation}
We define the power spectrum of  tensor fluctuations as 
\begin{equation}
    \left< h_{\bm{k}}^{(\sigma)} h_{\bm{k}'}^{(\sigma)}\right> = \frac{2\pi^{2}}{k^{3}}P^{(\sigma)}(k)
                               \delta^{(3)}(\bm{k}+\bm{k}') \ .
\end{equation}
Let us divide  the tensor fluctuations into the two parts. 
The one comes from vacuum fluctuations and the other comes from the gauge field.  
Since they are uncorrelated to each other, we can write the tensor power spectrum as  the sum 
\begin{equation}
    P^{(\sigma)}(k) = P^{(\sigma)}_{v}(k)+P^{(\sigma)}_{s}(k) \label{power}\ ,
\end{equation}
where $P_{v}(k)=2\times( \frac{H_{I}}{\pi M_{pl}})^{2}$ . 
From  Eqs.(\ref{hmode})-(\ref{power}),   we can deduce 
\begin{eqnarray}
    P^{(\sigma)}_{s}(k) = \frac{k^{3}}{\pi^{2}M_{pl}^{4}a^{2}} 
                &&\ \int \frac{d^{3}p}{(2\pi)^3} \left( 1+(\hat{k}\cdot\hat{p})^{2} \right) 
                 \left( 1+(\hat{k} \cdot \widehat{\bm{k}-\bm{p}})^{2} \right)  \nonumber \\
      &&  \times\left| \int d\tau' a^3 (\tau' )  G_{k}(\tau,\tau')\mathcal{E}_{p}(\tau')
      \mathcal{E}_{|\bm{k}-\bm{p}|}(\tau') \right|^{2}\ ,
      \label{power5}
\end{eqnarray}
where we ignored the subdominant contribution of the magnetic field and used 
an identity 
$\left( \vec{e}^{(\sigma)}(\hat{k})\cdot \vec{e}^{(\rho)}(\hat{k}') \right)^{2}
=\frac{1}{4}\left( 1-\sigma\rho (\hat{k}\cdot\hat{k}') \right)^2$. 
 The Green's function $G_{k}(\tau,\tau')$ for  Eq.(\ref{hmode})  is given by 
\begin{eqnarray}
    G_{k}(\tau,\tau')\simeq&\ \frac{1}{k^{3}\tau\tau'}\left[\ k\tau'\cos(k\tau')-\sin(k\tau') \ \right] \quad
                    &\left(\ \left| k\tau \right| \ll 1 \ \right)\ ,\nonumber \\
               \simeq& -\frac{\tau'^{2}}{3\tau} \hspace{4cm}
               &\left(\ \left| k\tau \right|, \ \left| k\tau' \right| \ll 1 \ \right)\label{green}\ .
\end{eqnarray}
Substituting  Eqs.(\ref{ele}) and (\ref{green}) into  Eq.(\ref{power5}) and using the new variables $\vec{q}\equiv \frac{\vec{p}}{k}$, $\vec{q'}\equiv \frac{\vec{p}-\vec{k}}{k}$ and $z \equiv -k\tau$, we get 
\begin{eqnarray}
    P_{s}(k) &=& 2\times P^{(\sigma)}_{s}(k)\nonumber \\
    &=&\frac{1}{36\pi^{7}} \left( \frac{H_{I}}{M_{pl}} \right)^{4} 
     \ \int d^{3}q \left( 1+(\hat{k}\cdot\hat{q})^{2} \right) \left( 1+(\hat{k} \cdot \hat{q'})^{2} \right) 
     \nonumber\\ 
    && \hspace{2.5cm}
    \times\left|\int dz' q^{-\nu+1/2}q'^{-\nu+1/2} \frac{2^{2\nu-1} \Gamma^{2}(\nu+\frac{1}{2})}{z'^{2\nu-3}}\right|^{2}
    \ ,
\end{eqnarray}
where we used the fact  that there is no polarization of gravitational waves. 
Note that $\nu$ depends on $z'$.
Since the gauge field becomes relevant as the source of the gravitational waves after 
 starting to grow, 
we consider the region  
\begin{equation}
    \left| \tau_{in} \right| > \frac{1}{p}\ ,\ \frac{1}{\left| \vec{p}-\vec{k} \right|} \ , \label{qin} 
\end{equation}
where $\tau_{in}$ is the time when the gauge field starts to grow.
Multiplying it by k, we get 
\begin{equation}
    \quad \frac{1}{q_{in}} > \frac{1}{q}\ ,\ \frac{1}{q'} \quad,
\end{equation}
where $q_{in} \equiv \left| k\tau_{in} \right|^{-1}$ represents the 
infrared cut off of the momentum integral.
Since the time integral is carried out in the range where the source and the Green's function has been 
on super horizon scales, 
we have the relation
\begin{eqnarray}
     \left| \tau \right| \ < \ \left| \tau' \right| \ < \ 
            \frac{1}{p}\ ,\ \frac{1}{\left| \vec{p}-\vec{k} \right|} \ , 
            \frac{1}{k}\ , \ \left| \tau_{in} \right| \quad .
\end{eqnarray}
From Eq.(\ref{qin}), it reads 
\begin{equation}
    \left| \tau \right| \ < \ \left| \tau' \right| \ < \ 
                 \frac{1}{p}\ ,\ \frac{1}{\left| \vec{p}-\vec{k} \right|}  \ ,\ \frac{1}{k} \quad.
\end{equation}
Multiplying $k$, we obtain 
\begin{equation}
      z < z' < \rm{min}\left[ \frac{1}{q}\ ,\ \frac{1}{q'} \ ,\ 1 \right] \quad .
\end{equation}
The dominant contribution to the momentum integral comes from  $q \sim q_{in}$ or $q' \sim q_{in}$\ .
Then $q\sim 1$ or $q' \sim 1$, namely,  the upper limit of the time integral becomes 
\begin{equation}
    \rm{min}\left[ \frac{1}{q}\ ,\ \frac{1}{q'}\ ,\ 1 \right] \sim 1 \ .
\end{equation}
Therefore, we obtain 
\begin{eqnarray}
    P_{s}(k) = 
    \frac{2}{9\pi^{6}} \left( \frac{H_{I}}{M_{pl}} \right)^{4} 
    \left|\int^{1}_{z} dz' \frac{2^{2\nu-1} \Gamma^{2}(\nu+\frac{1}{2})}{\sqrt{\nu-2}}
      \left( \frac{1}{q_{in}} \right)^{\nu-2} z'^{-2\nu+3} \right|^{2} \label{power2}\ .
\end{eqnarray}
To calculate the integration (\ref{power2}), we must know the function $\nu(z')$.
Taking a look at  Eq.(\ref{nupote}), we see an explicit form of the inflaton potential $V(\phi)$ is necessary. 
Here, we consider a monomial inflaton potential, namely $V(\phi)\propto \phi^{l}$. 
Then  we obtain
\begin{equation}
    \nu= \frac{l c M_{pl}}{\phi} \label{phi}\ .
\end{equation}
In this case,  Eq.(\ref{alpha2}) leads to 
\begin{equation}
    \alpha = -\frac{1}{2l}\frac{\phi^{2}}{M_{pl}^{2}}+D \ .
\end{equation}
We set $\alpha=0$, when $\phi=\frac{clM_{pl}}{2}$ which is the moment the gauge field starts to grow (see Eq.(\ref{condition2})).
Thus, we obtain the relation
\begin{equation}
    \alpha = \frac{1}{2l}\left[ \frac{l^{2}c^{2}}{4}-\frac{\phi^{2}}{M_{pl}^{2}} \right] \label{alpha}\ .
\end{equation}
It represents the number of e-foldings counting from the beginning of growth of the gauge field. 
For convenience, we defined several key moments in Fig.\ref{fig1}. 
Using these variables, we can deduce the following expression  
\begin{eqnarray}
 q_{in}\equiv \frac{p_{in}}{k}=e^{-(N_{k}-N_{in})}\ ,
    \quad z'  \equiv -k\tau'=e^{-(N-N_{k})} \ ,
    \quad z  \equiv -k\tau=e^{-(N_{e}-N_{k})} \ ,
\end{eqnarray}
where $N$ corresponds to the variable $z'$. 
Since $\alpha=N-N_{in}$ , we get 
\begin{equation}
    \alpha=-\ln(z')-\ln(q_{in}) \ .
\end{equation}
Similarly, we have the relation 
\begin{equation}
    N_{tot}\equiv N_{e}-N_{in}=-\ln(z)-\ln(q_{in}) \ .
\end{equation}
Consequently, we have
\begin{equation}
    \alpha =N_{tot}+\ln(z)-\ln(z') \label{alpha5}\ .
\end{equation}
Substituting Eq.(\ref{alpha5}) into  Eq.(\ref{alpha}), we obtain 
\begin{equation}
    \phi=M_{pl}\sqrt{\frac{l^{2}c^{2}}{4}-2l \left( N_{tot}+\ln(z)-\ln(z') \right)} \label{phi2} \quad .
\end{equation}
Finally, from  Eqs.(\ref{phi}) and (\ref{phi2}), we get 
\begin{equation}
    \nu=\frac{lc}{\sqrt{\frac{l^{2}c^{2}}{4}-2l\left( N_{tot}+\ln(z)-\ln(z') \right)}} \quad .
\end{equation}
Using this relation, we can carry out the integral (\ref{power2}). 
Before that, we need  to determine $N_{tot}$ . 
Notice that the scalar field at the end of inflation $\phi_{min}$  is determined by the 
breakdown of the slow roll condition
$\epsilon=\frac{M_{pl}^{2}}{2}\left( \frac{V_{,\phi}}{V} \right)^{2}=1$.
This gives $\phi_{min}^2 = M_{pl}^2 l^2 /2$ .
On the other hand, from  Eq.(\ref{phi2}), we find  the minimum value of the inflaton is obtained by setting $z=z'$ as 
\begin{equation}
    \phi_{min}=M_{pl}\sqrt{\frac{l^{2}c^{2}}{4}-2l N_{tot}} \quad .
\end{equation}
Thus, we can determine $N_{tot}$ as 
\begin{equation}
    N_{tot}=\frac{l}{8}\left( c^{2}-2 \right) \ .
\end{equation}

\begin{figure}[H]
\begin{center}
\includegraphics[width=11cm]{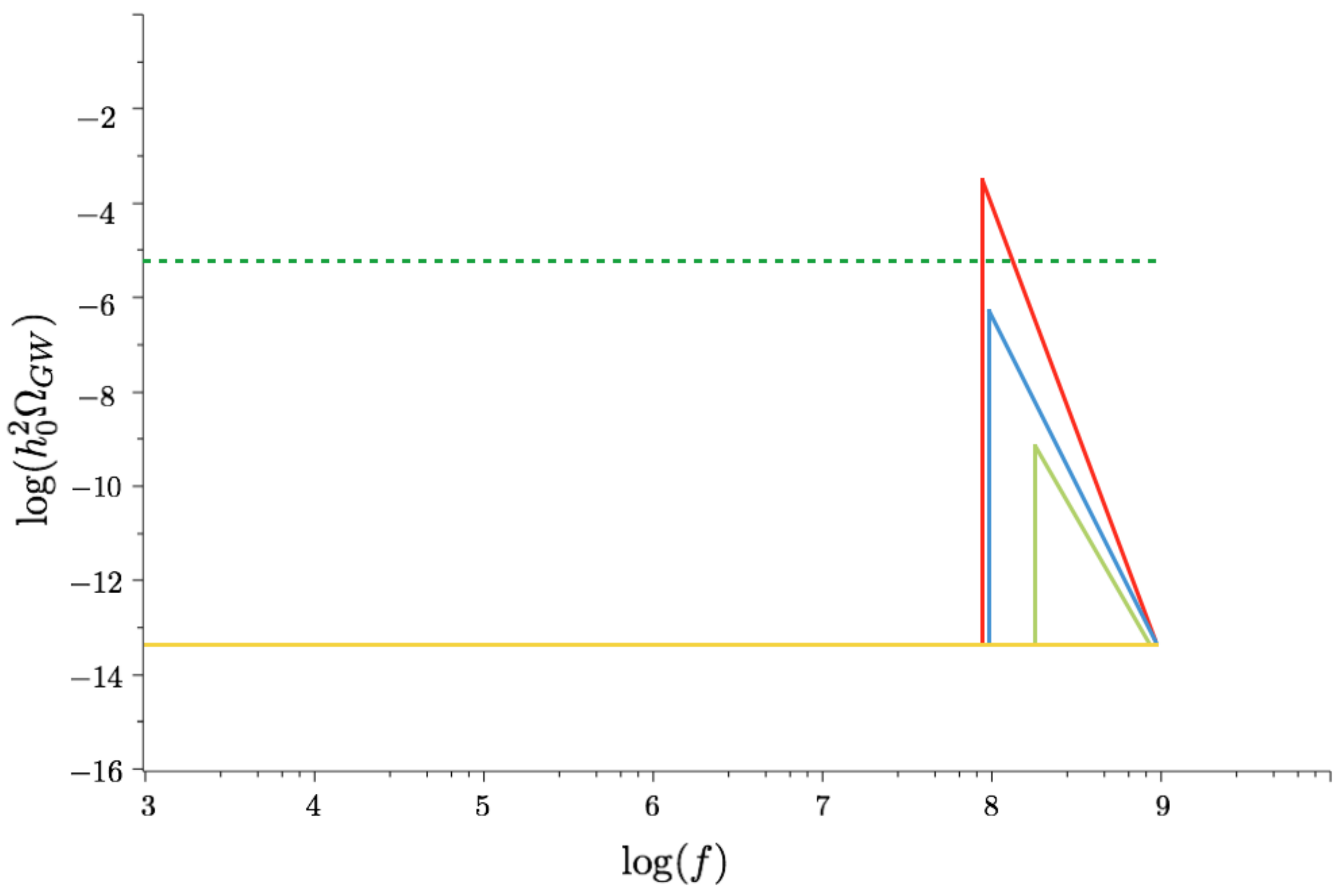}
\end{center}
\caption{In the case of $l=2$, the spectrum of gravitational waves are depicted. 
The red peak represents the case of $c=3.6$, the blue and green one correspond to $c=3.45$ and 
$c=3.3$, respectively. 
}
\label{l2}
\end{figure}
\begin{figure}[H]
\begin{center}
\includegraphics[width=11cm]{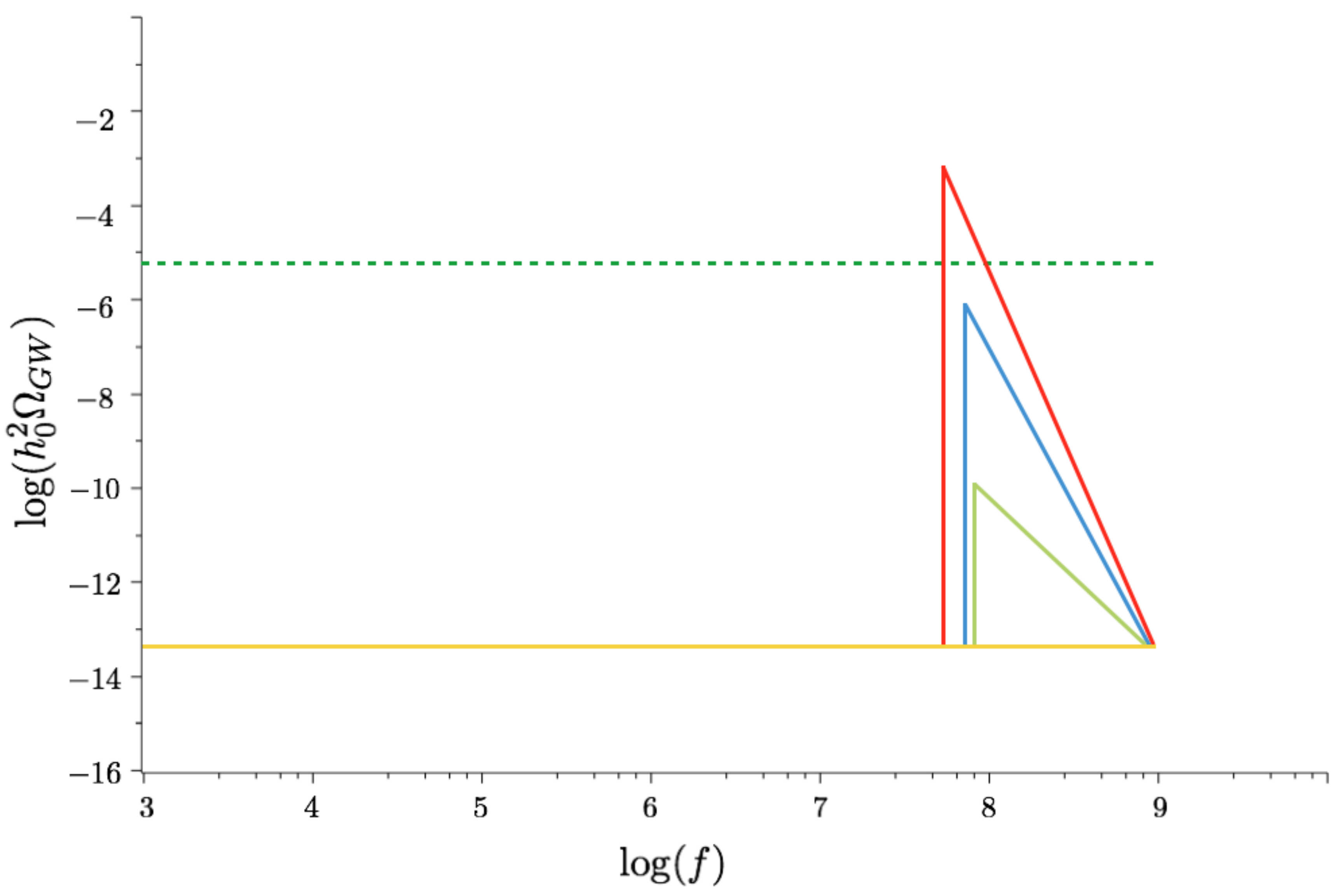}
\end{center}
\caption{In the case of $l=4$, the spectrum of gravitational waves are depicted. 
The red peak represents the case of $c=3.1$, the blue and green one correspond to $c=3$ and 
$c=2.9$ respectively.}
\label{l4}
\end{figure}
\begin{figure}[H]
\begin{center}
\includegraphics[width=11cm]{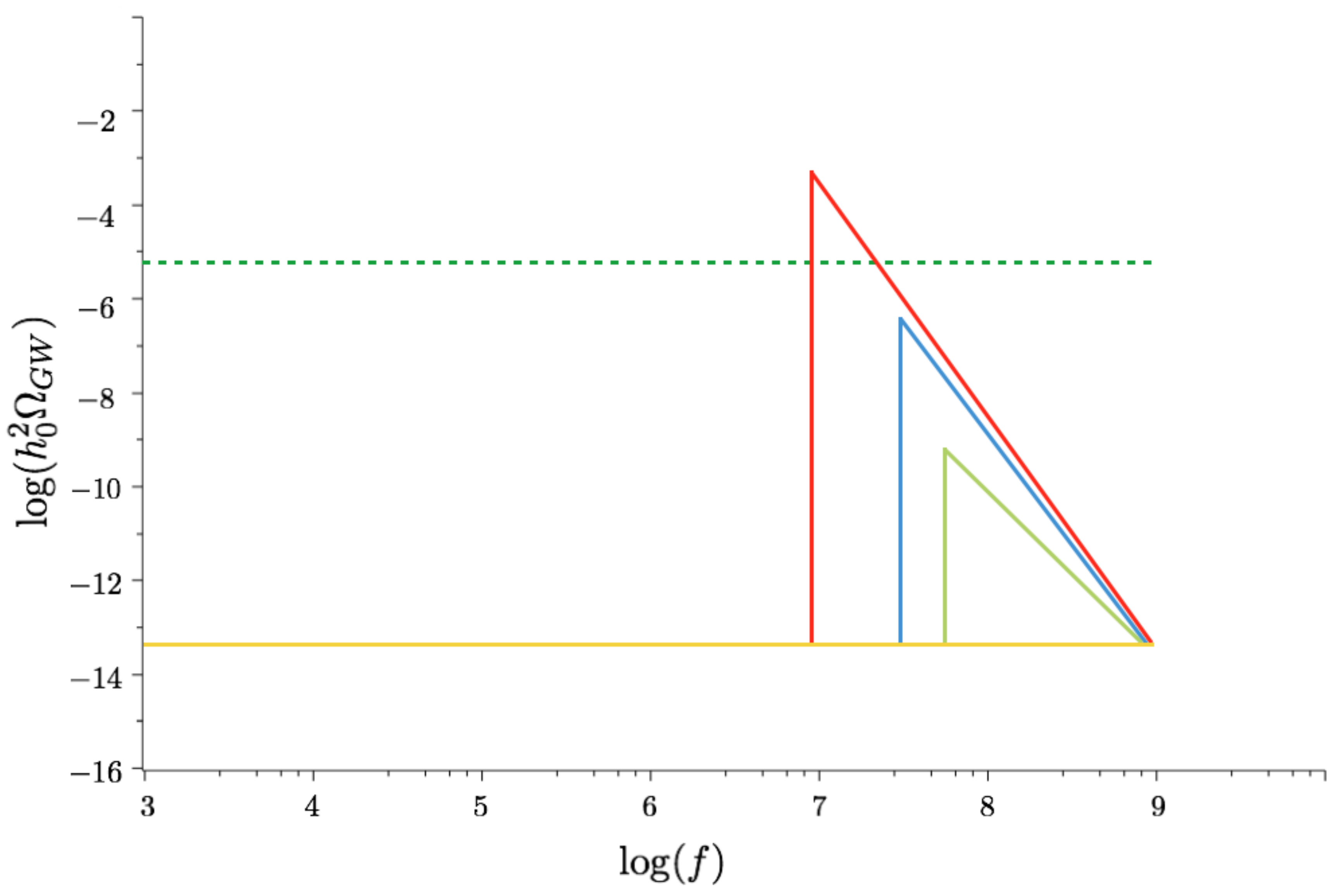}
\end{center}
\caption{In the case of $l=6$, the spectrum of gravitational waves  are depicted. 
The red peak represents the case of $c=2.9$, the blue and green one correspond to $c=2.8$ and 
$c=2.7$ respectively.}
\label{l6}
\end{figure}

We are now in a position to calculate the integral (\ref{power2}) for several values of  $l$ and $c$. 
The  results are shown in  Figs.\ref{l2}-\ref{l6}, where we took
the Hubble constant to be $H_{I}=10^{-5}M_{pl}$ . 
In Figs.\ref{l2}-\ref{l6},  the spectrum of the energy density parameter $h_{0}^{2}\Omega_{GW}$ 
($h_{0}$ is the dimensionless Hubble parameter) of gravitational waves 
are depicted against the frequency of gravitational waves. 
The yellow line represents conventional vacuum tensor fluctuations constrained by CMB observations. 
The green dotted line is an upper bound to the primordial gravitational waves from big-bang nucleosynthesis. 
In the case of $l=4,6$ (Figs.\ref{l4},\ref{l6}), 
we have assumed that the inflaton potential on CMB scales is different from 
that near the end of inflation. It is assumed that they are connected with each other smoothly. 
From the results, we see that gravitational waves can be copiously produced by the gauge field 
in the $e^{c \phi/M_{pl}}FF$ model.
There exists a peak in the spectrum of the  density parameter because
the scales which have exited the horizon earlier are more enhanced than the one exited later. 
Note that the amplitude can exceed the bound of the nucleosynthesis depending on the parameters. 
As can be seen, gravitational waves can be produced for a short period. 
It is because that $\nu$ of Eq.($\ref{phi}$) grows in the late stage of inflation and 
it affects the integral ($\ref{power2}$) significantly. 
As to observations, several works to detect the gravitational waves in MHz$\sim$GHz frequency bands 
have been carried out \cite{Nishizawa:2007tn}. 
The current sensitivity for the amplitude of gravitational waves at the frequency 100MHz is $h_{c}\sim 10^{-13}$ 
\cite{Kuroda:2015owv}, where the characteristic strain is defined by $h_{c}\equiv \sqrt{\frac{3}{2}}
\frac{H_{0}\Omega_{GW}^{1/2}}{\pi f}$. 
On the other hand, our model can reach  $10^{-26}\sim 10^{-27}$ 
in the frequency bands 10MHz$\sim$100MHz. 
Although the current sensitivity is too low to detect the signal, 
recently new methods are suggested and under development \cite{newexp}. 
Since scalar fluctuations are also produced by the same mechanism, 
one may worry about primordial black holes. 
In fact, the mass of primordial black holes depend on the reheating temperature and 
we need consider each models individually. 
However, as long as the reheating temperature is not so low, primordial black holes with mass range produced at the end of inflation have no severe constraints because they evaporate well before the big-bang nucleosynthesis.

\section{Conclusion}
We  claimed short-term anisotropic inflation is universal in the sense that it appears in models with a broad class
of potential and gauge kinetic functions.
As a concrete demonstration, we studied anisotropic inflation with an exponential 
type gauge kinetic function (\ref{exp}) which is ubiquitous in fundamental theory.
For example, it is obtained by dimensional  reduction from higher dimensional fundamental theory. 
In the $e^{c \phi/M_{pl}}FF$ model, it turned out that the gauge field starts to grow in the late stage of 
inflation and consequently anisotropic inflation commences for a generic parameter $c\sim \mathcal{O}(1)$. 

Remarkably, short-term anisotropic inflation provides an interesting phenomenology.
Indeed, we have shown that the gauge field during short-term anisotropic inflation produces copious 
gravitational waves in the frequency band 10MHz $\sim$ 100MHz. 
The amplitude of the gravitational waves can reach about $10^{-26}\sim 10^{-27}$. 
Such gravitational waves could be detectable in near future \cite{newexp} and 
it enable us to probe fundamental theory. 
We note that our model is quite different from other models which predict the existence of 
 high-frequency primordial gravitational waves \cite{Giovannini:1999bh} 
in that the spectrum of our model has a peak and a red-tilt  rather than a blue-tilt. 

There are several applications to be considered. 
It is interesting to investigate the creation of cosmological magnetic fields in the present model. 
Since the gauge field begins to grow in the last stage of inflation, 
we may have a different scenario for  primordial magnetic fields \cite{Kanno:2009ei}. 
It is also possible to extend the present model to non-abelian cases \cite{Murata:2011wv} 
or other inflaton  potentials.

\acknowledgements
AI would like to thank Ippei Obata and Arata Aoki for useful conversations.
This work was supported by  JSPS KAKENHI Grant Number 25400251,
 MEXT  KAKENHI Grant Number 26104708, and MEXT KAKENHI Grant Number 15H05895.

\end{document}